\documentclass[journal]{IEEEtran}

\usepackage{amsfonts}
\usepackage{amsmath}
\usepackage{amssymb}
\usepackage{graphicx}
\usepackage{mathtools}
\usepackage{adjustbox}
\usepackage{hyperref}
\hypersetup{colorlinks,allcolors=black}
\usepackage{cite} 

\usepackage{bbm}  

\usepackage{xcolor} 

\usepackage{soul}  

\usepackage{amsthm} 

\usepackage[normalem]{ulem}
\newtheorem{theorem}{Theorem}

\newtheorem{example}{Example}

\begin{document}
\title{A Tree Pruning Technique for Decoding Complexity Reduction of Polar Codes and PAC Codes}
\author{Mohsen~Moradi\textsuperscript{\href{https://orcid.org/0000-0001-7026-0682}{\includegraphics[scale=0.06]{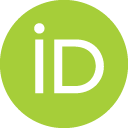}}},
Amir~Mozammel\textsuperscript{\href{https://orcid.org/0000-0003-3474-9530}{ \includegraphics[scale=0.06]{figs/ORCID}}}
\thanks{The authors are with the Department of Electrical-Electronics Engineering, Bilkent University, Ankara TR-06800, Turkey (e-mail: moradi@ee.bilkent.edu.tr, a.mozammel@ee.bilkent.edu.tr).}%
}

\maketitle
\begin{abstract}
Sorting operation is one of the main bottlenecks for the successive-cancellation list (SCL) decoding. This paper introduces an improvement to the SCL decoding for polar and pre-transformed polar codes that reduces the number of sorting operations without degrading the code's error-correction performance. In an SCL decoding with an optimum metric function we show that, on average, the correct branch's bit-metric value must be equal to the bit-channel capacity, and on the other hand, the average bit-metric value of a wrong branch can be at most zero.
This implies that a wrong path's partial path metric value deviates from the bit-channel capacity's partial summation. For relatively reliable bit-channels, the bit metric for a wrong branch becomes very large negative number, which enables us to detect and prune such paths. We prove that, for a threshold lower than the bit-channel cutoff rate, the probability of pruning the correct path decreases exponentially by the given threshold. Based on these findings, we presented a pruning technique, and the experimental results demonstrate a substantial decrease in the amount of sorting procedures required for SCL decoding. In the stack algorithm, a similar technique is used to significantly reduce the average number of paths in the stack.

\end{abstract}
\begin{IEEEkeywords}
PAC codes, polar codes, SCL decoding, sequential decoding, polarized channel, stack decoding, pruning.
\end{IEEEkeywords}

\section{Introduction}
\IEEEPARstart{}{}
Polar codes are capable of achieving the capacity of memoryless symmetric channels \cite{arikan2009channel}, and as an advance over successive cancellation (SC) decoding, SC list (SCL) decoding enables polar codes to perform well throughout block lengths ranging from short to moderate with a high-rate outer code \cite{tal2015list}.
During each decoding step of an SCL decoding with a list size of $L$, $L$ decoding paths are explored at each level of the decoding process.
The SCL decoder doubles the number of possible decoding paths for each information bit and then performs a sorting technique to eliminate all but the $L$ most probable decoding possibilities.

Fano algorithm is a well-known sequential decoding algorithm that lends itself nicely to hardware implementations \cite[ch. 6.4]{jacobsPrinciples}.
The algorithm searches just the most probable path specified by its metric function. 
Suppose the metric function's value starts to drop. 
In that case, the sequential decoder recognizes that the current path is incorrect and attempts to explore another path until it reaches the end of the decoding tree.
The Fano algorithm has a backtracking feature that makes it to visit some nodes several times before determining the correct path.
Benefiting from the technique of tracking only one path of the Fano decoding algorithm, we propose an improvement to the SCL decoding algorithm that eliminates those paths that have a metric value below a certain threshold. 
We adopt the averaging technique of sequential decoding bit-channel metrics of \cite{moradi2021sequential} to prove that the correct path's metric value of SCL decoding should be equal to the summation of the bit-channel capacities.
By polarization technique, polar coding uses the provided noiseless bit-channels to transmit data through them. 
The bit metric of a wrong branch on average is a negative number, and we prove that for a noiseless bit channel, the bit metric of a wrong branch is equal to $-\infty$.
On the other hand, the bit-metric of the correct branch on average is equal to the bit-channel capacity (which is a positive number), and for a noiseless channel, we prove that it should be equal to 1.
We utilize this to exclude those paths whose bit-metric values are large negative numbers.
For the purposes of our simulations, we discard paths that have bit-metric values less than a threshold $m_T$, and the results indicate that, in the majority of decoding levels, we can keep the number of paths to be ordered below the list size $L$ with no degradation in error-correction performance.
We prove that the probability of discarding the correct branch at each level decays exponentially with $m_T$.
Our results indicate that when a list decoding is used, our suggested algorithm requires almost no sorting operations while decoding a $(1024, 512)$ polar code at high signal-to-noise ratio (SNR) values with a similar error-correction performance.

Based on identifying the best path, in \cite{chen2013reduced} and \cite{chen2015reduce}, pruning techniques are used after the sorting operations of the SCL decoding algorithm (sorting $2L$ branches is necessary for identifying the best path). 
Although this approach adds a computational overhead to the list pruning operation, it can substantially lower the total computational complexity.
We show that our proposed technique is an improvement to these works, and our main goal is to avoid the sorting operation as much as possible.
Numerical examples indicate that in addition to reducing the number of sorting operations, our proposed approach outperforms the technique provided in \cite{chen2013reduced} in terms of average unit computations without incurring additional computational overhead.

While using a sequential decoding algorithm, the error-correction performance of the newly introduced polarization-adjusted convolutional (PAC) codes approaches the theoretical bounds with a very low computational complexity at high SNR values \cite{arikan2019sequential, moradi2020performance, moradi2021sequential, moradi2021monte, mozammel2021hardware}.
Additionally, practical results demonstrate that when the list size $L$ is somewhat large, equivalent error-correction performance can be attained using list decoding algorithm \cite{yao2021list}. 
We also demonstrate the effectiveness of our suggested algorithm on the list decoding of PAC codes.
Note that avoiding the sorting procedure becomes increasingly important when the list size is large.

One main drawback of the Fano algorithm is that it may visit various nodes of the decoding tree many times during its execution.
As another widely known sequential decoding technique, stack decoding algorithm would only visit each node at most once by keeping the examined paths in a sorted stack \cite{zigangirov1966some}, \cite{jelinek1969fast}. 
Both algorithms eventually choose identical paths through the tree, and the set of nodes seen by the Fano and stack algorithms are identical \cite{geist1970algorithmic}. 
The primary concern is that the stack size may grow excessively large, which restricts its broad adoption. 
Similar to the SCL decoding, we introduce an improvement to the stack algorithm that makes use of the bit-metric values and results in a stack that contains, on average, much less paths in it compared to the conventional stack algorithm. 
The latency of the stack decoding algorithm is heavily reliant on the number of paths in the stack and the numerical results indicate that our proposed technique results in much less paths in the stack for a PAC$(128, 64)$ code while maintaining the same error-correction performance.

All of the vectors and matrices in the paper are over the binary Galois field $GF(2).$
Random variables are indicated by capital letters and their realizations by lowercase letters.
Matrices are denoted using bold capital letters.
Vectors are denoted using bold lowercase letters, and given a vector $\mathbf{x} = (x_1, x_2, ..., x_N) \in GF(2)^N$, the subvector $(x_1, x_2, ..., x_i)$ is denoted by $\mathbf{x}^i$, and the subvector $(x_i, ..., x_j)$ is denoted by $\mathbf{x}_i^j$ for $i\leq j$. 
For each subset of indices $\mathcal{A} \subset \{1, 2, ..., N\}$, $\mathcal{A}^c$ represents the complement of $\mathcal{A}$, and $\mathbf{x}_{\mathcal{A}}$ denotes the subvector $(x_i : i\in \mathcal{A})$.

The remainder of this article is structured as follows.
Sections \ref{sec: polar code} and \ref{sec: PAC codes} discuss briefly the encoding of polar and PAC codes. 
Section \ref{sec: path metric} explains and analyzes the optimal path metric of SCL decoding and addresses the probability that the bit-metric of a correct path falls below a certain threshold.
In Section \ref{sec: prposed alg}, we develop a pruning technique to improving the SCL decoding algorithm.
Section \ref{sec: stack algorithm} reviews the stack algorithm.
Section \ref{sec: reliable_stack} introduces an improvement to the stack algorithm that benefits from the bit-metric values.
Section \ref{sec: dynamic} then suggests a dynamic threshold for the pruning technique.
Finally, Section \ref{sec: conclusion} provides a succinct overview of this work.

\section{Polar codes}\label{sec: polar code}

\begin{figure}[htbp]
\centering
	\includegraphics [width = \columnwidth]{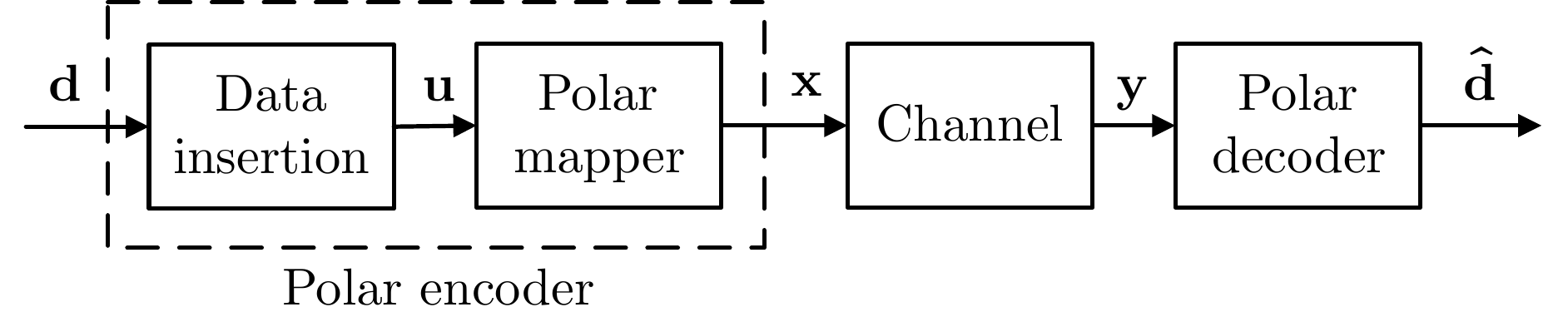}
	\caption{Polar coding scheme.} 
	\label{fig: polar_encoding}
\end{figure}

A polar coding scheme is shown in Fig. \ref{fig: polar_encoding}. 
In general, three parameters $(N, K, \mathcal{A})$ can be used to specify a polar code, where $N = 2^n$ for $n \geq 1$ is the codeword length, $K$ is the data word length which can be any integer number from $1$ to $N$, 
and set $\mathcal{A}$, known as the data index set, is a subset of $\{1, 2, \cdots , N\}$ of size $|\mathcal{A}| = K$. 
For an arbitrary codewrod length of $N$, the encoding operation may be expressed in terms of the generator matrix $\mathbf{F}^{\otimes n}$ defined recursively as
\begin{equation}
\mathbf{F}^{\otimes {n}} =
    \begin{bmatrix} 
    \mathbf{F}^{\otimes {n-1}} & \mathbf{0}\\ 
    \mathbf{F}^{\otimes {n-1}} & \mathbf{F}^{\otimes {n-1}} 
    \end{bmatrix},
\end{equation}
where
\begin{equation}
    \mathbf{F}^{\otimes {1}} = \mathbf{F} =
    \begin{bmatrix} 1 & 0\\ 1 & 1 \end{bmatrix}.
\end{equation}

The data insertion in the block diagram is accomplished by inserting the bits of the data vector $\mathbf{d} = (d_1, \cdots, d_K)$ into the bits of a data-carrier vector $\mathbf{u} = (u_1, u_2, \cdots, u_N)$ as $\mathbf{u}_{\mathcal{A}^c} = \mathbf{0}$ (we assume a memoryless symmetric channel)
and $\mathbf{u}_{\mathcal{A}} = \mathbf{d}$, resulting in an encoding rate of $R = K/N$.
Then the polar transformation $\mathbf{F}^{\otimes n}$ is used to map the vector $\mathbf{u}$ into the vector $\mathbf{x}$ as $\mathbf{x} = \mathbf{u}\mathbf{F}^{\otimes n}$.
This vector is transmitted through the channel, and at the receiver side, the data vector estimates $\hat{\mathbf{d}}$ are recovered from the channel output $\mathbf{y}$ by using a polar decoder such as SC or SCL decoding algorithms. 

The overall channel transition probability $W_N(\mathbf{y}|\mathbf{u})$ between the input vector $\mathbf{u} = (u_1, u_2, \cdots, u_N)$ and the vector of channel outputs $\mathbf{y} = (y_1, y_2, \cdots, y_N)$ for a binary-input discrete memoryless channel (B-DMC) is defined as
\begin{equation}
    W_N(\mathbf{y}|\mathbf{u}) \triangleq
    \Pi_{i=1}^{N}W(y_i|x_i) = 
    \Pi_{i=1}^{N}W(y_i|(\mathbf{u}\mathbf{F}^{\otimes n})_i),
\end{equation}
where $W(y_i|x_i)$ is the channel transition probability.
After polarization, the $i$th subchannel transition probability ($i$th bit-channel) with the input-output pair as $(u_i; \mathbf{y},\mathbf{u}^{i-1})$ is obtained as
\begin{equation}
    W_N^{(i)}(\mathbf{y},\mathbf{u}^{i-1}|u_i) =
    \frac{1}{2^{N-1}}\sum_{\mathbf{u}_{i+1}^{N}}W_N(\mathbf{y}|\mathbf{u}).
\end{equation}

Given a discrete memoryless channel,
\begin{equation}
    I(W) \triangleq 
    \sum_{y\in \mathcal{Y}}
    \sum_{x\in \mathcal{X}}
    q(x)W(y|x)\log 
    \frac{W(y|x)}{\sum_{x^{'}\in \mathcal{X}}q(x^{'})W(y|x^{'})}
\end{equation}
denotes the average mutual information (symmetric capacity) between the inputs and outputs where $q(x)$ is the probability distribution on the inputs.

Another fundamental channel parameter is the Gallager's function of a channel $W$ which is defined as 
\begin{equation}\label{eq: error exponent}
    E_0(\rho, W) = - \log_2 \sum_{y \in \mathcal{Y}} \left[ \sum_{x \in \mathcal{X}} q(x) W(y|x)^{\frac{1}{1+\rho}}\right] ^{1+\rho} ,
\end{equation}
where $\rho \geq 0$ and
the channel cutoff rate is defined as $E_0(1,W)$ \cite[p. 138]{gallager1968information}.
For the $i$th bit-channel, we denote its bit-channel capacity by $I(W_N^{(i)})$ and its bit-channel cutoff rate by $E_0(1,W_N^{(i)})$.
When $N$ goes to infinity, $E_0(1,W_N^{(i)})$ converges to $I(W_N^{(i)})$ \cite{arikan2015origin}.
Moreover, Gallager's function polarization is established in \cite{alsan2014polarization}.
Based on channel polarization, asymptotically, almost $NI(W)$ of the bit-channels would have their bit-channel capacities close to one (noiseless bit channels), about $N(1-I(W))$ would have their bit-channel capacities close to 0, and the fraction of intermediate channels vanishes.

\section{PAC codes}\label{sec: PAC codes}
Polar codes' performance for small block lengths falls significantly short of theoretical boundaries due to its poor weight distribution.
Polar codes' weight distribution can be improved by utilizing a  pre-transformation \cite{li2019pre}. 
At small block lengths, a novel pre-transformed polar coding method named PAC codes significantly outperforms conventional polar codes by benefiting from a convolutional pre-transformation \cite{arikan2019sequential, moradi2020performance}. 

Four parameters $(N, K, \mathcal{A}, \mathbf{T})$ can be used to specify a PAC code. Similar to polar codes, $N, K$ and $\mathcal{A}$ are the codeword length, data word length, and data index set, respectively.
$\mathbf{T}$ is an upper-triangular Toeplitz matrix formed using a generator polynomial $\mathbf{c}(t) = c_mt^{m} + \cdots + c_1t + c_0$, with $c_0 = c_m = 1$ and can be represented as
\begin{equation}
\mathbf{T} = 
\begin{bmatrix}
 c_0    & c_1    &  c_2   & \cdots & c_m    & 0      & \cdots & 0      \\
 0      & c_0    & c_1    & c_2    & \cdots & c_m    &        & \vdots \\
 0      & 0      & c_0    & c_1    & \ddots & \cdots & c_m    & \vdots \\
 \vdots & 0      & \ddots & \ddots & \ddots & \ddots &        & \vdots \\
 \vdots &        & \ddots & \ddots & \ddots & \ddots & \ddots & \vdots \\
 \vdots &        &        & \ddots & 0      & c_0    & c_1    & c_2    \\
 \vdots &        &        &        & 0      & 0      & c_0    & c_1    \\
 \vdots & \cdots & \cdots & \cdots & \cdots & 0      & 0      & c_0    
\end{bmatrix}.
\end{equation}

\begin{figure}[htbp]
\centering
	\includegraphics [width = \columnwidth]{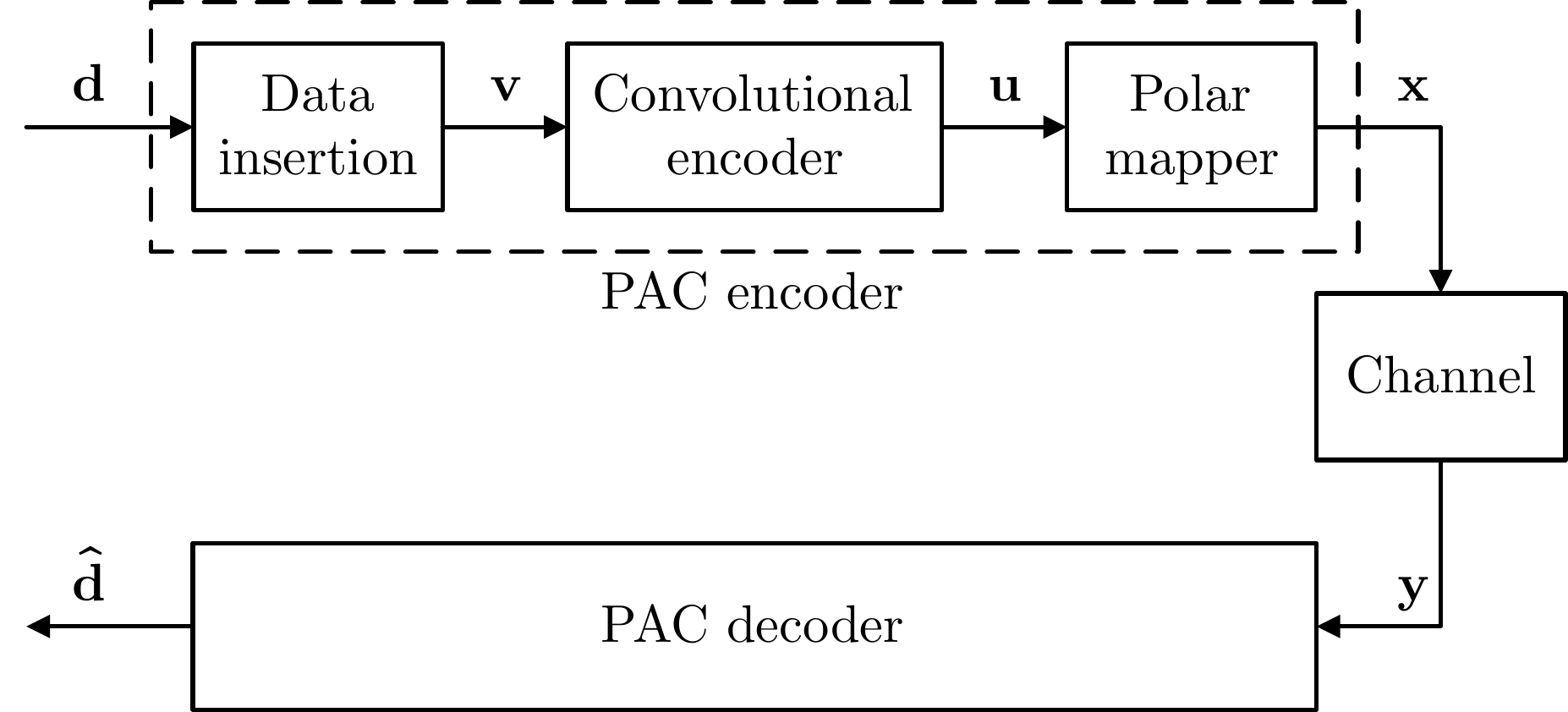}
	\caption{PAC coding scheme.} 
	\label{fig: PAC_encoding}
\end{figure}
Fig. \ref{fig: PAC_encoding} illustrates the PAC encoding scheme in block diagram form. 
Three components comprise the PAC encoding process: data insertion, convolutional encoder, and polar mapper.
Similar to the polar codes, the data insertion block embeds the data vector $\mathbf{d}$ into a data-carrier vector $\mathbf{v}$.
This procedure, alongside with the selection of the index set $\mathcal{A}$, is referred to as \textit{rate-profiling}. 
Polar rate-profiling and Reed-Muller (RM) rate-profiling are two known construction methods for obtaining this set $\mathcal{A}$ \cite{arikan2019sequential}. 
A novel rate-profiling technique for polarized channels is introduced in \cite{moradi2021monte}.
In the convolutional encoder block, vector $\mathbf{u}$ can be obtained as $\mathbf{u} = \mathbf{v}\mathbf{T}$ for the upper-triangular Toeplitz matrix $\mathbf{T}$.
This generates the vector $\mathbf{u}$, each bit of which is a linear combination of at most $(m+1)$ bits of the vector $\mathbf{v}$ calculated by the convolution operation.
Finally, the vector $\mathbf{u}$ is encoded into $\mathbf{x}$ as $\mathbf{x} = \mathbf{u}\mathbf{F}^{\otimes n}$.

For a specified matrix $\mathbf{T}$ and the index set $\mathcal{A}$, we use PAC$(N,K)$ notation to show a PAC code with $(N, K, \mathcal{A}, \mathbf{T})$ parameters.

\section{Path metric function for SCL decoder}\label{sec: path metric}
In the SCL decoding, whenever the $i$th bit is an information bit, the decoding tree considers both bifurcated paths, and the decoder chooses $L$ out of all possible options. 
To accomplish this, the decoder requires a path metric function capable of calculating the likelihoods throughout the decoding process to determine the most likely path based on that function. 
For a given channel output $\mathbf{y}$, the optimal metric function in the terms of error probability is to choose $\hat{\mathbf{u}}^i$ as the sequence $\mathbf{u}^i$ for which the value of $P(\mathbf{u}^{i}|\mathbf{y})$ is maximized in the $i$th level of decoding.
Note that the SC decoder has a likelihood value for frozen bits as well and that given a wrong path, the decoder may recommend a nonzero value for a frozen bit.
This may assist the SCL decoder in determining that this path is one of the wrong paths, so the metric function should consider both the frozen and information bit likelihoods. 

By using Bayes' rule, $P(\mathbf{u}^{i}|\mathbf{y})$ may be expressed as 
\begin{equation}\label{eq: Bayes rule_MAP}
    P(\mathbf{u}^{i}|\mathbf{y}) = \frac{P(\mathbf{y}|\mathbf{u}^{i})}{P(\mathbf{y})}P(\mathbf{u}^{i}).
\end{equation}

Notice that the $P(\mathbf{u}^{i})$ term has no influence on the maximality owing to the symmetric channels and having the same path lengths in SCL decoding and can be avoided in calculating the path metric. 
Because the monotonically increasing $\log$ function maintains maximality, the optimum metric also maximizes 
\begin{equation}
    \Phi(\mathbf{u}^i;\mathbf{y}) \triangleq \log_2 \frac{P(\mathbf{y}|\mathbf{u}^{i})}{P(\mathbf{y})},
\end{equation}
which is called the partial path metric of SCL decoding.
Therefore, in the $i$th level of the decoding tree, the bit-metric function can be obtained as
\begin{equation}\label{eq: Listbitmetric}
    \begin{split}
        \phi(u_i ;\mathbf{y},\mathbf{u}^{i-1})
        &  \triangleq \Phi(\mathbf{u}^{i};\mathbf{y}) - \Phi(\mathbf{u}^{i-1};\mathbf{y})\\
        & = \log_2 
        \frac{P(\mathbf{y}|\mathbf{u}^{i})}{P(\mathbf{y})}
         -
        \log_2 \frac{P(\mathbf{y}|\mathbf{u}^{i-1})}{P(\mathbf{y})}\\
        & =   \log_2 \frac{P(\mathbf{y},\mathbf{u}^{i-1} | u_{i})}{P(\mathbf{y},\mathbf{u}^{i-1})} .
    \end{split}
\end{equation}

In {\eqref{eq: Listbitmetric}}, $\mathbf{y}$ is the physical channel output and $\mathbf{u}^{i-1}$ are the correct (true) inputs that are provided by a genie.
The last equality of {\eqref{eq: Listbitmetric}} is due to the memorylessness of the input distribution.
For any given output $(\mathbf{y},\mathbf{u}^{i-1})$ of a bit-channel $W_N^{(i)}$, we have
$\phi(u_i ;\mathbf{y},\mathbf{u}^{i-1}) \geq \phi(\tilde{u}_i ;\mathbf{y},\mathbf{u}^{i-1})$ if and only if $P(\mathbf{y},\mathbf{u}^{i-1} | u_{i}) \geq P(\mathbf{y},\mathbf{u}^{i-1} | \tilde{u}_{i})$. 
That is, when dealing with a bit-channel $W_N^{(i)}$, the bit metric $\phi$ is calculated according to the local ML rule.

For a binary input channel with a uniform input distribution the bit metric can be represented as 
\begin{equation}\label{eq: bitmetricformula}
\begin{split}
    & \phi(u_{i} ;\mathbf{y},\mathbf{u}^{i-1}) = \log_2 \left(\frac{P(\mathbf{y},\mathbf{u}^{i-1}|u_{i} )}{P(\mathbf{y},\mathbf{u}^{i-1})}\right)
    \\
    & = \log_2 \left(\frac{P(\mathbf{y},\mathbf{u}^{i-1}|u_{i} )}{\frac{1}{2} \left[P(\mathbf{y},\mathbf{u}^{i-1}|u_{i}) +P(\mathbf{y},\mathbf{u}^{i-1}|u_{i} \oplus 1)  \right] }\right)
    \\
    & = 1 - \log_2
    \left( 1 + \frac{P(\mathbf{y},\mathbf{u}^{i-1}|u_{i} \oplus 1)}{P(\mathbf{y},\mathbf{u}^{i-1}|u_{i})}
    \right) \\
    & = 1 - \log_2
    \left( 1 +
    2^{-\log_2\left(\frac{P(\mathbf{y},\mathbf{u}^{i-1}|u_{i})}{P(\mathbf{y},\mathbf{u}^{i-1}|u_{i} \oplus 1)} \right)}
    \right)  \\
    & = 1 - \log_2 \left(
    1 + 2^{- L_i \mathord{\cdot}(-1)^{u_i}}
    \right),
\end{split}
\end{equation}
where
\begin{equation}
    L_i = \log_2 \left( \dfrac{P(\mathbf{y},\mathbf{u}^{i-1}|u_i = 0 )}{P(\mathbf{y},\mathbf{u}^{i-1}|u_i = 1 )}
    \right).
\end{equation}
Therefore, for a $0$ branch or a frozen bit, the bit metric would be $ 1 - \log_2 \left(1 + 2^{- L_i} \right)$ and for a $1$ branch the bit metric is $1 - \log_2 \left(1 + 2^{L_i} \right)$.

In this part, in order to make the analysis feasible, we introduce randomness into the model shown in Fig \ref{fig: polar_encoding} by establishing a probabilistic model.
We define an ensemble of random variables denoted by $(\mathbf{U}, \mathbf{X}, \mathbf{Y})$, which corresponds to the variables of the system denoted by $(\mathbf{u}, \mathbf{x}, \mathbf{y})$ in Fig. \ref{fig: polar_encoding}. 
Due to the system design, the joint distribution 
$P_{\mathbf{U}, \mathbf{X}, \mathbf{Y}}(\mathbf{u}, \mathbf{x}, \mathbf{y})$
takes the form 
$q_{\mathbf{U}}(\mathbf{u})
P_{\mathbf{X}|\mathbf{U}}(\mathbf{x|\mathbf{u}})
P_{\mathbf{Y}|\mathbf{X}}(\mathbf{y|\mathbf{x}})$,
where 
$P_{\mathbf{X}|\mathbf{U}}(\mathbf{x|\mathbf{u}}) = \mathbbm{1}(\{\mathbf{x}=\mathbf{u}\mathbf{F}^{\otimes n} \})$
and $P_{\mathbf{Y}|\mathbf{X}}(\mathbf{y|\mathbf{x}}) =
\Pi_{i=1}^{N}W(y_i|x_i)$,
and $\mathbbm{1}(.)$ represents the indicator function. 
In order to get a distribution $P_{\mathbf{U}, \mathbf{X}, \mathbf{Y}}$ that is computationally feasible, we assume that the binary random variables $U_i$ are independent and uniformly distributed.

\begin{figure}[htbp]
\centering
	\includegraphics [width = .6\columnwidth]{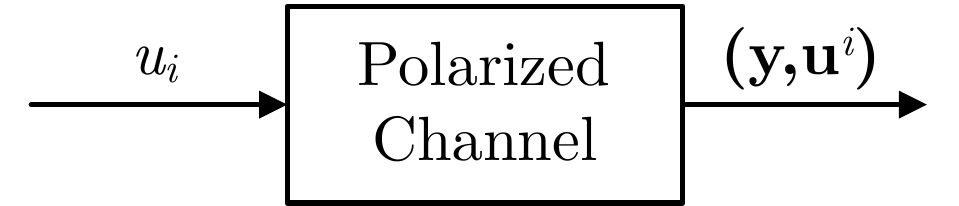}
	\caption{Bit-channel scheme.} 
	\label{fig: bitChannel}
\end{figure}

Due to the channel polarization, we may further simplify our analysis of the system model of Fig. \ref{fig: polar_encoding} by examining the system model of Fig. \ref{fig: bitChannel}.
We present an ensemble of random variables denoted by $(U_i, \mathbf{Y}, \mathbf{U}^{i-1})$ that correspond to the variables of the system denoted by $(u_i, \mathbf{y}, \mathbf{u}^{i-1})$ in Fig. \ref{fig: bitChannel}.
The joint distribution 
$P_{U_i, \mathbf{Y}, \mathbf{U}^{i-1}}(u_i, \mathbf{y}, \mathbf{u}^{i-1})$ may be represented by the distribution $q_{U_i}(u_i)
P_{(\mathbf{Y}, \mathbf{U}^{i-1})|U_i}(\mathbf{y}, \mathbf{u}^{i-1} | u_i)$.

In a decoding algorithm, the metric function determines which direction through the decoding tree should be chosen. 
As a matter of intuitive understanding, the bit-metric function value must always improve for every $u_i$ bit on the correct path in order to detect the correct data vector $\mathbf{u}$.
It is impossible to place a measure on this point of view for a certain code. 
Studying an ensemble of codes with the specified channel and code length may make the problem easier to understand. 
The drift of the bit metric is defined as the expectation 
$\mathbb{E} \left[\phi(U_i ;\mathbf{Y},\mathbf{U}^{i-1})\right]$ where the expectation is computed with respect to a distribution that depends on whether $u_i$ is a node on the correct path or not. 
It is not possible to determine the drift of the path metric based on the precise code information.
In its stead, the drift is estimated using the joint ensemble described above.

In the case of the ensemble of bit channels with output-input  $(\mathbf{y},\mathbf{u}^{i-1} ; u_i)$ pair ($\mathbf{y}$ represents the physical channel output, whereas the vector $\mathbf{u}^{i-1}$ is the correct inputs provided by a genie), the expectation of \eqref{eq: Listbitmetric} is
\begin{equation}\label{eq: av_correctpath_SCL}
    \begin{split}
        & \mathbb{E}_{U_i,\left(\mathbf{Y},\mathbf{U}^{i-1} \right)} \left[\phi(U_i ;\mathbf{Y},\mathbf{U}^{i-1})\right]\\
        & = \sum_{u_i}q(u_i) \sum_{(\mathbf{y},\mathbf{u}^{i-1})} P(\mathbf{y},\mathbf{u}^{i-1}|u_i) \phi(u_i ;\mathbf{y},\mathbf{u}^{i-1})
        \\
       & = \sum_{u_i}\sum_{(\mathbf{y},\mathbf{u}^{i-1})}q(u_i)P(\mathbf{y},\mathbf{u}^{i-1}|u_i) \left[ \log_2 \left(\frac{P(\mathbf{y},\mathbf{u}^{i-1}|u_i)}{P(\mathbf{y},\mathbf{u}^{i-1})}\right)  \right] \\
       & = I(W_N^{(i)}) ,
    \end{split}
\end{equation}
where $I(W_N^{(i)})$ denotes the symmetric capacity of the bit-channel $W_N^{(i)}$. 
As a consequence, the heuristic result is that for the bit channels the average metric increases all the time for the correct branches.

Suppose now that $\Tilde{u}_i$ is a wrong bit with a metric of $ \phi (\Tilde{u}_i ; \mathbf{y}, \mathbf{u}^{i-1}) $ at level $i$. 
By averaging this bit-metric value over the correct and incorrect branches, we have
\begin{equation}
    \label{eq: av_wrongpath_SCL}
    \begin{split}
        & \mathbb{E}_{U_{i}, \Tilde{U}_i,\left(\mathbf{Y},\mathbf{U}^{i-1} \right) } \left[\phi(\Tilde{U}_i ;\mathbf{Y},\mathbf{U}^{i-1})\right]\\
        & = \sum_{\Tilde{u}_i} q(\Tilde{u}_i) \sum_{u_{i}}q(u_{i}) \sum_{(\mathbf{y},\mathbf{u}^{i-1})} P(\mathbf{y},\mathbf{u}^{i-1}|u_{i})
        \phi(\Tilde{u}_i ;\mathbf{y},\mathbf{u}^{i-1})
        \\
        & =\sum_{\Tilde{u}_i} q(\Tilde{u}_i) \sum_{(\mathbf{y},\mathbf{u}^{i-1})} P(\mathbf{y},\mathbf{u}^{i-1}) \phi(\Tilde{u}_i ;\mathbf{y},\mathbf{u}^{i-1})
        \\
        & = \sum_{\Tilde{u}_i}\sum_{(\mathbf{y},\mathbf{u}^{i-1})}q(\Tilde{u}_i)P(\mathbf{y},\mathbf{u}^{i-1})
        \left[ \log_2 \left(\frac{P(\mathbf{y},\mathbf{u}^{i-1}|\Tilde{u}_i)}{P(\mathbf{y},\mathbf{u}^{i-1})}\right)  \right]
        \\
        & \leq \frac{1}{\ln{(2)}} \sum_{\Tilde{u}_i}\sum_{(\mathbf{y},\mathbf{u}^{i-1})}q(\Tilde{u}_i)P(\mathbf{y},\mathbf{u}^{i-1}) \left[ \frac{P(\mathbf{y},\mathbf{u}^{i-1}|\Tilde{u}_i)}{P(\mathbf{y},\mathbf{u}^{i-1})} - 1  \right]  \\
         & =  0.
    \end{split}
\end{equation}

It should be noted that we used the $\ln(x) \leq x-1$ inequality for $x >0$ to obtain \eqref{eq: av_wrongpath_SCL}. 
Because the distribution of the channel output $(\mathbf{y},\mathbf{u}^{i-1})$ is dependent on $u_i$, we were required to carry out an averaging operation on the weighting of the correct input $u_i$ as well.
For a genie-aided decoder, the expectation stated above means that, on average, for each wrong branch, the bit metric is a negative or $0$ value.
This means that a path with couple of wrong branches, on average, has a low probability and possibly the decoder can eliminate the path from the list of possible paths.

The above discussion is about the behaviour of the bit-channel metric on average. 
Polarization attempts to obtain $K$ noiseless bit channels out of all $N$ bit channels. 
Lets assume that we achieve to this goal and the $i$th bit channel is noiseless. 
In this case, if we have transmitted $u_i$ bit and $\tilde{u}_i = u_i\oplus1$, we have
\begin{equation}
\begin{split}
& \phi(u_i;\textbf{y},{\textbf{u}}^{i-1}) 
= \log_2\frac{\text{P}(\textbf{y},{\textbf{u}}^{i-1}|{u}_i)}{\text{P}(\textbf{y},{\textbf{u}}^{i-1})}\\
& = \log_2\frac{\text{P}(\textbf{y},{\textbf{u}}^{i-1}|{u}_i)}{\frac{1}{2}\text{P}(\textbf{y},{\textbf{u}}^{i-1}|{u}_i) + \frac{1}{2}\text{P}(\textbf{y},{\textbf{u}}^{i-1}|{\tilde{u}}_i)}\\
& = \log_2\frac{1}{\frac{1}{2} + 0} = 1.
\end{split}
\end{equation}
This is in consist with \eqref{eq: av_correctpath_SCL} as for the noiseless bit channel, its bit-channel mutual information is 1.
Also for the $i$th wrong branch, we have its bit-metric function as
\begin{equation}
\begin{split}
& \phi(\tilde{u}_i;\textbf{y},{\textbf{u}}^{i-1}) 
= \log_2\frac{\text{P}(\textbf{y},{\textbf{u}}^{i-1}|{\tilde{u}}_i)}{\text{P}(\textbf{y},{\textbf{u}}^{i-1})}\\
& = \log_2\frac{0}{0+\frac{1}{2}} = -\infty.
\end{split}
\end{equation}
As a result of this finding, any pathways that include the wrong branch of a noiseless bit channel can never belong to a correct path. 
Also, from \eqref{eq: av_wrongpath_SCL} we know that the bit metric of the wrong branch should be a negative number on average. 
Based on this discussion, if the branch metric is a big negative number, the branch is the wrong one, and in the SCL decoding, there is no need to consider this branch in the $L$ survivor branches, and we can consider this bit channel as a noiseless bit channel. 
As explained, ideally (when the noise is small and code length is large enough), the bit metric of the wrong branch should be $-\infty$ ( the only way to get a noiseless bit channel is if the code block length is increased to infinity).
In the following subsection, we will prove that if the path was correct up to step $i-1$, the probability that the bit-metric for the correct bit $u_i$ falls below a certain threshold will exponentially go to zero with that threshold. 
Hence, this gives rise to the notion of pruning paths with very small metrics.

\subsection{Probability of error}\label{sec: prb error}
We make the assumption that the list size is infinite and that the failure to get correct decoding is solely due to the pruning strategy that we are proposing (this is the case for the stack decoding). 
We fail to successfully decode a received vector if the bit metric of the correct path is less than the threshold $m_T$ at the $i$th level for $i$ from $1$ to $N$.
Responding to this issue is equivalent to determining an upper bound for
\begin{equation}
    P\left\{\phi(U_i;\textbf{Y},{\textbf{U}}^{i-1}) \leq m_T \right\},
\end{equation}
where $m_T < I(W_N^{(i)})$ (in our simulations we use $m_T < 0$). 
To accomplish this, we use the Chernoff bound:

\textbf{Chernoff bound.}
For a random variable $X$, if $m < \mathbb{E}[X]$, we have
\begin{equation}
    P\{ X \leq m\} \leq 2^{-sm}g(s); ~~~~ s < 0,
\end{equation}
where $g(s)$ is the moment-generating function (MGF) of $X$.
\hfill\IEEEQEDhere 

\begin{theorem}
For the random variable $\phi(U_i;\textbf{Y},{\textbf{U}}^{i-1})$ and constant threshold $m_T$ we have
\begin{equation}
    P\left\{\phi(U_i;\textbf{Y},{\textbf{U}}^{i-1}) \leq m_T \right\} \leq
    2^{m_T - E_0(1,W_N^{(i)})},
\end{equation}
where $m_T < I(W_N^{(i)})$.
\end{theorem}

\begin{proof}
We found the mean of the random variable $\phi(U_i)  \triangleq \phi(U_i;\textbf{Y},{\textbf{U}}^{i-1})$ as $I(W_N^{(i)})$, and in the following, we derive an upper bound for its MGF when $-1<s<0$. 
\begin{equation}
\begin{split}
    & g(s) \triangleq \mathbb{E}[2^{s\phi(U_i)}] = 
    \mathbb{E}\left[2^{s\log_2 \left(\frac{P(\mathbf{Y},\mathbf{U}^{i-1}|U_i)}{P(\mathbf{Y},\mathbf{U}^{i-1})}\right)  } \right] \\
    & = \mathbb{E}\left[ \left(\frac{P(\mathbf{Y} , \mathbf{U}^{i-1} | U_i)}{P(\mathbf{Y} , \mathbf{U}^{i-1})} \right)^{s} \right]\\
    & = \sum_{u_i}q(u_i)\sum_{(\mathbf{y},\mathbf{u}^{i-1})} P(\mathbf{y} , \mathbf{u}^{i-1} | u_i)
    \left(\frac{P(\mathbf{y} , \mathbf{u}^{i-1} | u_i)}{P(\mathbf{y} , \mathbf{u}^{i-1})} \right)^{s}\\
    & = \sum_{(\mathbf{y},\mathbf{u}^{i-1})} \underbrace{P(\mathbf{y},\mathbf{u}^{i-1})^{-s}}_a
    \underbrace{\sum_{u_i} q(u_i) P(\mathbf{y} , \mathbf{u}^{i-1} | u_i)^{1+s}}_b.
\end{split}   
\end{equation}

By defining $r = -s$ and considering $-1 < s < 0$ and using the H\"{o}lder's inequality (H) inequality:
\begin{equation}
    \sum ab \leq \left(\sum a^{\frac{1}{r}} \right)^{r}
    \left(\sum b^{\frac{1}{1-r}} \right)^{1-r},
\end{equation}
we have
\begin{equation}
\begin{split}
    & g(s) 
    \overset{\text{H}}{\le}
     \left[\underbrace{\sum_{(\mathbf{y},\mathbf{u}^{i-1})}P(\mathbf{y},\mathbf{u}^{i-1})}_\text{= 1} \right]^{r}\\
    &~~~~~~~~~ \left[\sum_{(\mathbf{y},\mathbf{u}^{i-1})} \left[\sum_{u_i} q(u_i) P(\mathbf{y} , \mathbf{u}^{i-1} | u_i)^{1-r} \right]^{\frac{1}{1-r}} \right]^{1-r}\\
    & =  2^{(1-r)\log_2 \left[\sum_{(\mathbf{y},\mathbf{u}^{i-1})} \left[\sum_{u_i} q(u_i) P(\mathbf{y} , \mathbf{u}^{i-1} | u_i)^{1-r} \right]^{\frac{1}{1-r}} \right]}\\
    & = 2^{ -(1-r)E_0(\frac{r}{1-r},W_N^{(i)})}
     = 2^{ -(1+s)E_0(\frac{-s}{1+s},W_N^{(i)})}.\\
\end{split}
\end{equation}

Therefore (to simplify), for $s = -1/2$ we have
\begin{equation}
\begin{split}
    &P\{ \phi(U_i) \leq m_T\} \leq 2^{-sm_T}g(s) \\
    &\leq 2^{-sm_T}2^{ -(1+s)E_0(\frac{-s}{1+s},W_N^{(i)})}\\
    & =2^{\frac{m_T - E_0(1,W_N^{(i)})}{2}}.
\end{split}
\end{equation}
This says that if the threshold is less than the bit-channel cutoff rate, the probability of pruning the correct path at the $i$th level of decoding goes exponentially to zero by $m_T$.
\end{proof}

Similar to the SCL decoding of polar codes, the bit- metric function for the SCL decoding of PAC codes can be represented as \eqref{eq: Listbitmetric} where the $i$th bit $u_i$ in the PAC code case is the input of the polar mapper.
One distinction from decoding polar codes is that for the $L$ paths of the list, $L$ vectors (auxiliary shift registers) with sizes equal to the constraint length of the convolutional code must be created to compute $u_i$ \cite{yao2021list}.
To acquire $L$ $u_i$s for \eqref{eq: bitmetricformula}, a $0$ should be placed into each shift register if it corresponds to a frozen bit or a zero branch; otherwise, a $1$ should be loaded into each shift register.


\section{Improving SCL decoding}\label{sec: prposed alg}
The primary downside of SCL decoding is that, given a list size of $L$, it is necessary to discover the $L$ greatest metric values out of $2L$ paths in each branching level of the decoding tree.
There are many algorithms for determining the $L$ greatest path metric values, and this stage of the SCL decoding process is often referred to as sorting \cite{tal2015list}. 
This section proposes an improvement to the SCL decoding algorithm (PSCL) that omits the sorting step for many of the decoding tree's branching levels.

\begin{figure}[htbp]
\centering
	\includegraphics [width = \columnwidth]{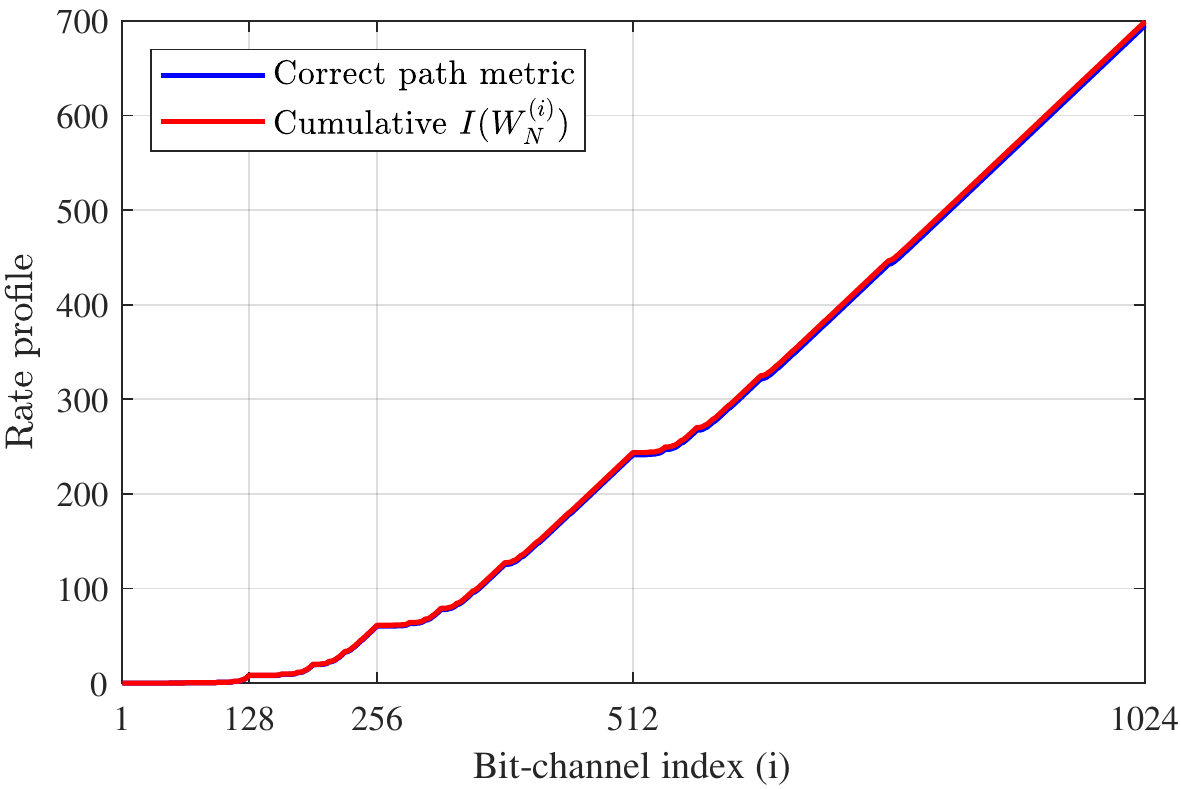}
	\caption{Comparison of the partial path metric corresponding to the correctly decoded codewords of list decoding of $(1024,512)$ polar code with the capacity rate profile over BI-AWGN at $2.5$ dB SNR.} 
	\label{ch2:fig: metric profile}
\end{figure}

As proved in the preceding section, the bit metric for the correct branch at the $i$th level of the decoding tree is $I(W_N^{(i)})$ on average, whereas the bit metric for an erroneous branch at this level would be less than $0$.
Fig. \ref{ch2:fig: metric profile} compares the average metric profiles for the correctly decoded codewords ($10^4$ codewords) to the capacity rate profile of a $(1024, 512)$ polar code over a binary-input additive white Gaussian noise (BI-AWGN) channel with an SNR of 2.5 dB. 
As this figure shows, for the noiseless bit channels (mostly the bit-channels that are at the end), the slope is 1 ($I(W_N^{(i)}) \approx 1$).
From this figure, it can be deduced that a wrong branch may be recognized from the correct branch when traversing through the decoding tree. 
To accomplish this, we propose modifying the SCL decoding algorithm as follows:

\begin{enumerate}
    \item At a branching level $i$, discard paths having a bit-metric $\phi(u_i;\mathbf{y}, \mathbf{u}^{i-1})$ less than the threshold $m_T$, where $m_T < I(W_N^{(i)})$.
    \item If the number of remaining paths exceeds the list size $L$, sort them and choose the $L$ ones with the highest metric values; otherwise, no sorting is required.
    \item Declare a decoding failure and end the decoding procedure if all $2L$ paths are rejected. 
    \item If $i = N$, select the path with the highest metric value.
\end{enumerate}

Note that the sorting of a typical SCL decoding technique is a function of $2L$, whereas the number of elements that may be sorted in our proposed algorithm can be much less than $2L$.

\begin{figure}[htbp]
\centering
	\includegraphics [width = \columnwidth]{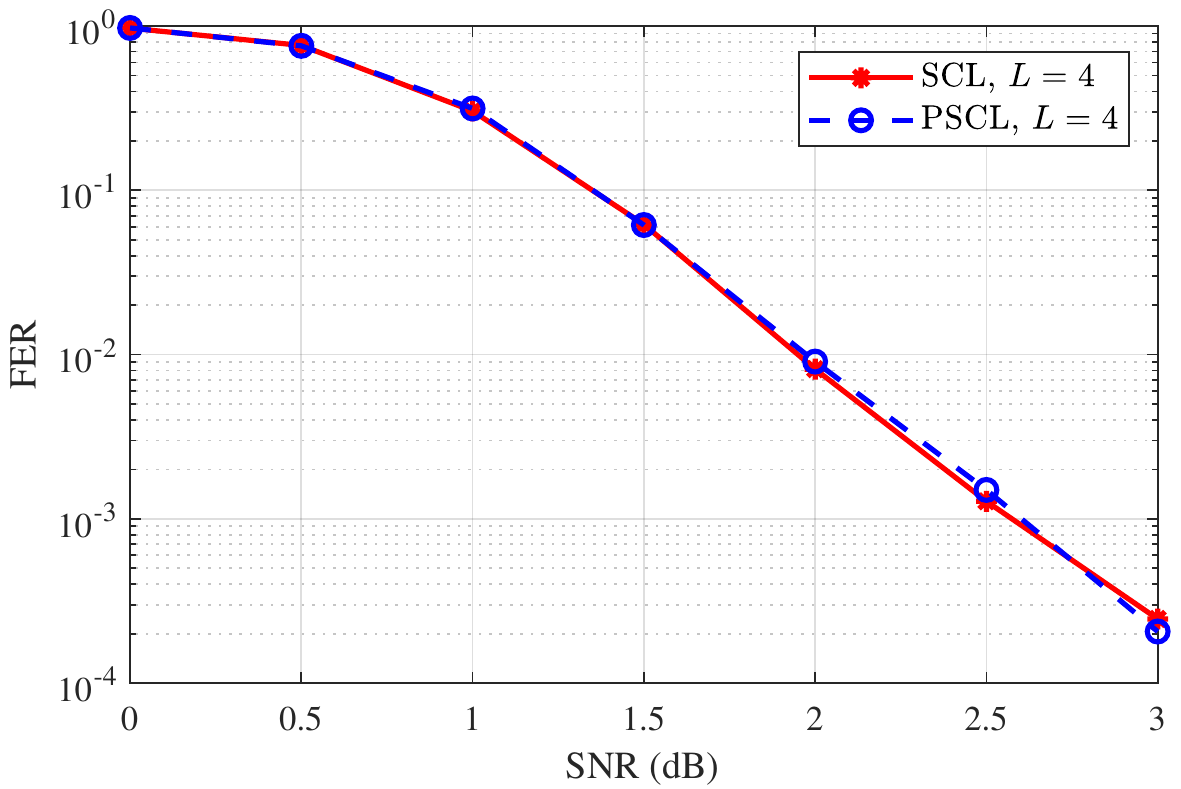}
	\caption{Comparison of the conventional SCL decoding with our proposed SCL decoding algorithm for a list size of $L = 4$ of a $(1024,512)$ polar code with $m_T = -5$.} 
	\label{ch2:fig: SCLvPSCLN1024}
\end{figure}

In Fig. \ref{ch2:fig: SCLvPSCLN1024}, the frame error rate (FER) performance of our proposed algorithm with the bit-metric threshold $m_T = -5$ for a $(1024, 512)$ polar code is compared to the performance of a conventional SCL decoding technique both with a list size of $L = 4$. 
As this figure demonstrates, there is no degradation in the FER performance.

\begin{table*}[htbp]
\footnotesize
\centering
\caption{Average number of executed sorting for decoding $(1024, 512)$ polar code.}
\renewcommand{\arraystretch}{1.2}
\begin{tabular}{cccccccc}
\hline
SNR {[}dB{]}          & 0.0    & 0.5    & 1.0    & 1.5    & 2.0    & 2.5    & 3.0    \\ \hline
\# of sorting ($L = 4$, $m_{T} = -5$) & 123.62 & 106.63 & 82.37 & 48.90 & 17.41 & 2.74 & 0.20 \\ \hline
\end{tabular}\label{ch2:table: polarN1024}
\end{table*}
 
The corresponding number of sorting operations performed in our proposed algorithm is listed in Table \ref{ch2:table: polarN1024}. 
Note that the conventional SCL decoding technique requires $510$ sorting operations to decode a $(1024, 512)$ polar code with a $L = 4$ list size. 
When the SNR is $3$ dB, our suggested decoding technique reduces the number of sorting operations to about zero without affecting the FER performance.

\begin{figure}[htbp]
\centering
	\includegraphics [width = \columnwidth]{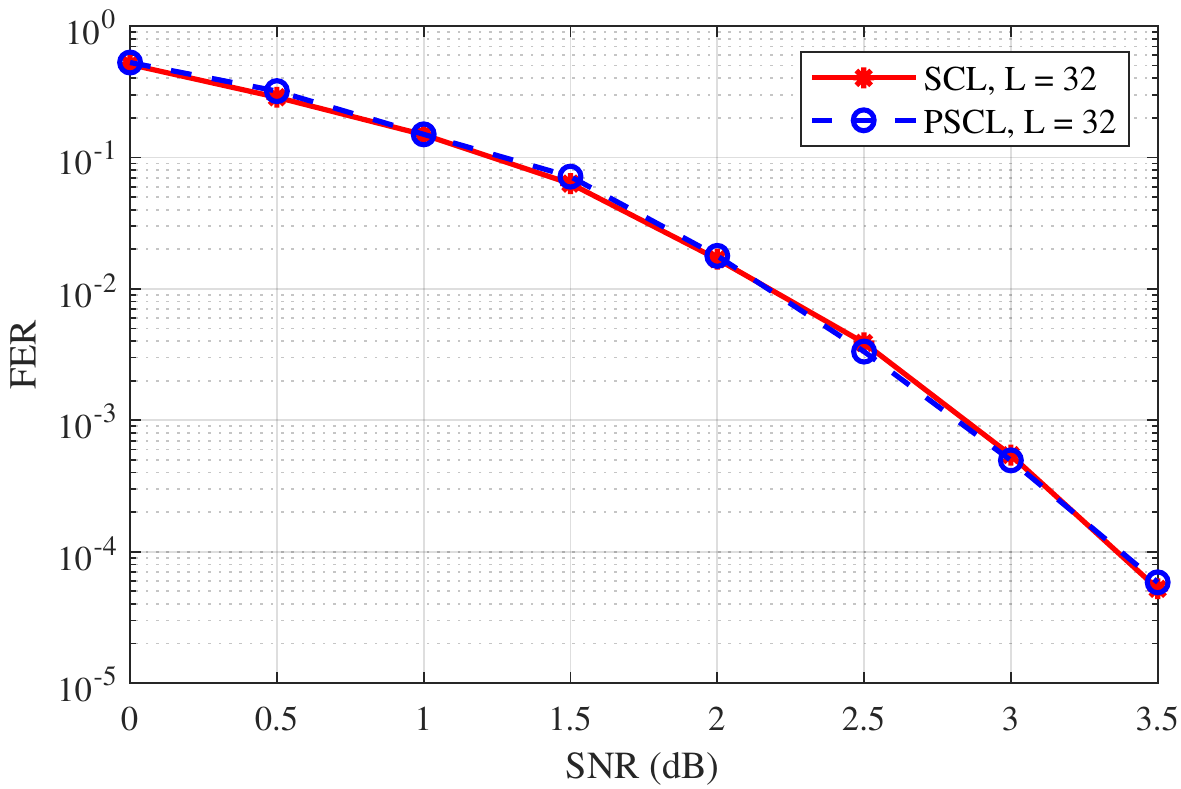}
	\caption{Comparison of the conventional list decoding with our proposed list decoding algorithm of a PAC$(128,64)$ code when $m_T =-10$.} 
	\label{ch2:fig: pac_SCLvPSCLN}
\end{figure}

Also, for a list size of $L = 32$, the FER performance of our proposed method for a PAC$(128,64)$ code constructed using RM rate profile is compared to the performance of a conventional SCL decoding technique in Fig. \ref{ch2:fig: pac_SCLvPSCLN}. 
As this figure illustrates, no loss in FER performance occurs.

\begin{table*}[htbp]
\footnotesize
\centering
\caption{Average number of executed sorting for decoding PAC$(128, 64)$ code.}
\renewcommand{\arraystretch}{1.2}
\begin{tabular}{ccccccccc}
\hline
SNR {[}dB{]}        & 0.0   & 0.5   & 1.0   & 1.5   & 2.0   & 2.5   & 3.0   & 3.5   \\ \hline
\# of sorting ($L = 32$, $m_T = -10$)  & 37.96 & 36.95 & 35.93 & 35.15 & 34.24 & 33.29 & 31.83 & 28.14 \\ \hline
\end{tabular}\label{ch2:table: pac}
\end{table*}

Table \ref{ch2:table: pac} lists the amount of sorting operations conducted by our proposed technique corresponding to the plot of Fig. \ref{ch2:fig: pac_SCLvPSCLN}. 
The conventional SCL decoding technique requires $59$ sorting operations to decode a PAC$(128, 64)$ code with a $L = 32$ list size.
As an instance, when the SNR is $3.5$ dB, our proposed decoding approach decreases the number of sorting operations to about $28$ without compromising the FER performance.
Note that rate profiling impacts the decoding complexity as well as the error-correction performance of the code \cite{moradi2021monte}.

A pruning technique is suggested in \cite{chen2013reduced} which is applied after the sorting step in the SCL algorithm and is based on the metric value of the best path.
Finding the best path at each level $i$ is costly, and our proposed algorithm avoids this by comparing the metric to the channel capacity profile, which is a priori information to the decoder. 
Similarly, in \cite{chen2015reduce} the pruning step is also after the sorting step of SCL decoding. 
An additional memory of size $L$ (stack) containing the information of the $L$ best paths that are already pruned (among those paths that are pruned, the paths in the stack have the highest metric values) is updated at every level of the decoding; 
the decoder maintains track of a maximum of $L$ survival paths and an additional $L$ paths in a memory called stack.
At each level $i$, a path metric will survive if its metric is better than the path metric of the stack top; otherwise, the path will be pruned. 
If the path is pruned and its metric is better than the metric of the worst path in the stack, the path will be inserted into the stack, and that worst path will be deleted out from the stack.
If the decoder is at the $i$th level and a path is pruned at level $j$ ($j < i$), in \cite{chen2015reduce} it is proposed to add its path metric with an upper bound on the metric to be able to use it at the level $i$.
We suggest updating the metric value of this pruned path by adding $\sum_{k = j+1}^i I(W_N^{(i)})$ which gives a better estimate of this path if it was not pruned and was decoded correctly up to the $i$th level (stack contains $L$ paths with such these updated metrics). 
The interesting thing about using this stack is that in \cite{chen2015reduce} they are trying to compare paths with different lengths. 
In a sequential decoding algorithm, the optimal way to compare the paths with different lengths is using a bias, and in \cite{moradi2021sequential} it is proved that the optimal bias should be the bit-channel capacities.
Sorting $2L$ paths in the SCL decoding process and sorting and updating the stack are costly operations. 
The primary purpose of our approach is to do as few sorting operations as possible. 
They can eliminate the sorting step in these works by just comparing the pathways to the bit-channel capacity profile (instead of the best path).

\begin{figure}[htbp]
\centering
	\includegraphics [width = \columnwidth]{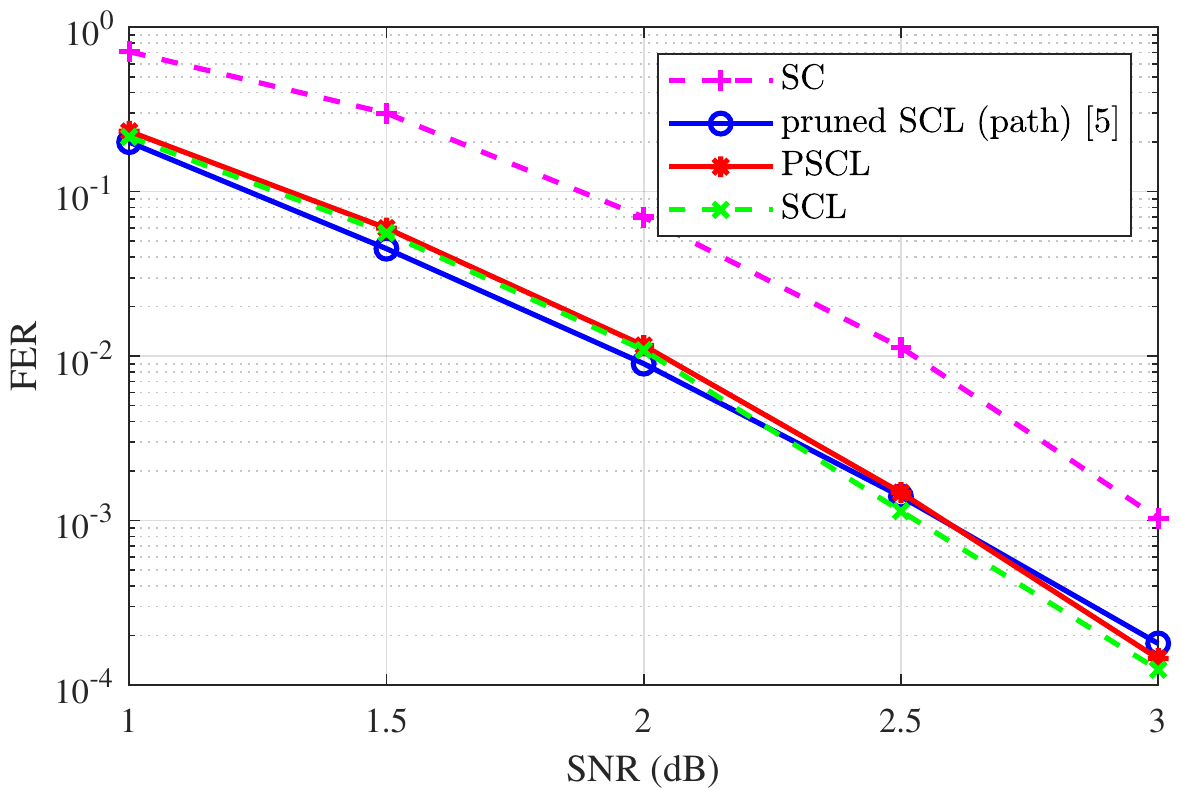}
	\caption{Comparison of the FER performance using different decoding techniques.}
	\label{fig: FER_pruned_BITvPATH}
\end{figure}

In Fig. \ref{fig: FER_pruned_BITvPATH}, we compare the FER performance of PSCL to the algorithm (pathwise pruned SCL) provided in \cite{chen2013reduced}. 
In this figure, the FER performance for SC and conventional SCL decoding algorithms are also plotted. 
This figure demonstrates that the error-correction performance of both techniques is close to that of the conventional SCL decoding algorithm.

\begin{figure}[htbp]
\centering
	\includegraphics [width = \columnwidth]{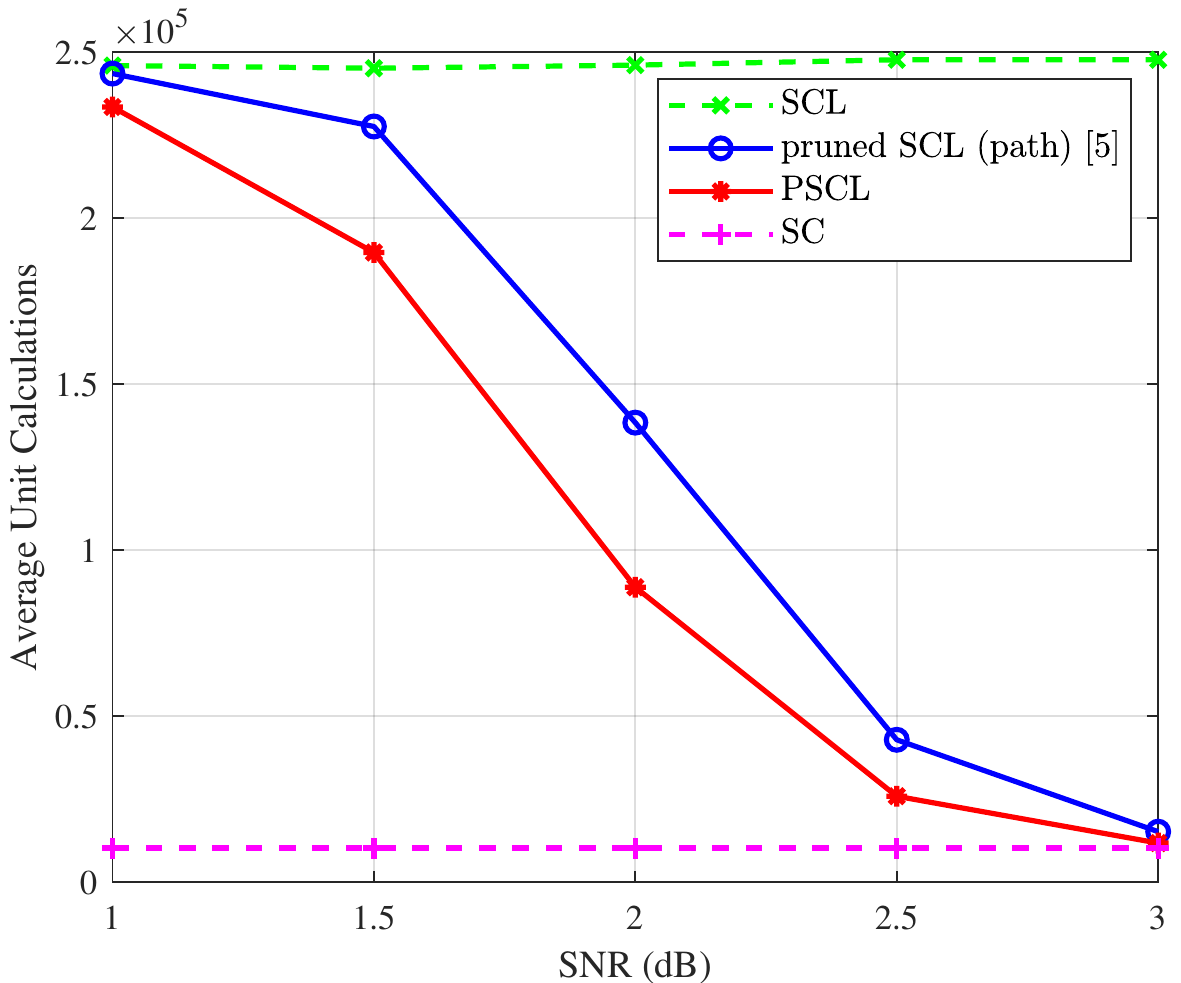}
	\caption{Comparison of the complexity using different decoding techniques.} 
	\label{fig: pruned_BITvPATH}
\end{figure}

In \cite{chen2013reduced}, the complexity of the pruning technique (in terms of average unit calculations) is determined by counting the number of $f$ and $g$-like operations. 
Fig. \ref{fig: pruned_BITvPATH} depicts a comparison of the complexity findings corresponding to the plots of the Fig. \ref{fig: FER_pruned_BITvPATH}.
$N\log_2(N) = 10240$ is the complexity of the SC decoding algorithm, which is achieved by both pruning techniques at high SNR levels. 
PSCL algorithm outperforms the pruning technique based on path metrics for all SNR values.
Note that the PSCL algorithm has the additional benefit of requiring less sorting.

\section{Stack Algorithm}\label{sec: stack algorithm}
The fundamental assumption of sequential decoding is that the decoder should evaluate just the most promising paths during the decoding process. 
Suppose a path to a node seems poor (based on its metric value). 
In this case, the decoder may reject all pathways originating from that node without incurring any significant performance loss relative to a maximum likelihood (ML) decoder.
The path metric function guides a decoder's search for the most promising path to investigate.
This naturally leads us to the stack algorithm \cite{zigangirov1966some, jelinek1969fast}, which uses a stack.
The following steps describe the stack algorithm for a PAC$(N, K)$ code:  

\begin{enumerate}
\item 
Calculate the path metric associated with the single-node path that includes just the origin node.
Stack the single-node path and its associated path metric value (initialize with $0$ metric). 
\item 
Calculate the path metric values of both immediate successor paths to the top path in the stack if the current node is an information bit; otherwise, calculate for the one immediate successor.
Delete the stack's top path. 
\item
After adding the remaining successor paths to the stack, reorder the stack in descending order according to the partial path metric values. 
\item
The algorithm terminates if the top path in the stack terminates at a leaf node in the decoding tree; otherwise, the algorithm proceeds to Step 2.
\end{enumerate}

The algorithm generates a stack of previously explored paths of varying lengths ordered descendingly by their partial path metric values. 
In this paper, we use
\begin{equation}\label{BranchMetric}
\begin{aligned}
\gamma(u_j;\textbf{y},{\textbf{u}}^{j-1}) 
&= \log_2\frac{P(\textbf{y},{\textbf{u}}^{j-1}|{u}_j)}{P(\textbf{y},{\textbf{u}}^{j-1})}- E_0(1,W_N^{(j)}),
\end{aligned}
\end{equation}
as the bit-metric function for the $j$th bit-channel, where $E_0(1,W_N^{(j)})$ is the bit-channel cutoff rate \cite{moradi2021sequential}.

This bit-metric function for sequential decoding is equivalent to the bit-metric function $\phi(u_j;\textbf{y},{\textbf{u}}^{j-1})$ for SCL decoding with the addition of the bias term $b_j \triangleq E_0(1,W_N^{(j)})$. 
In SCL decoding, all paths have the same length. 
However, in sequential decoding, this bias term enables the comparison of paths with different lengths. 
Therefore, the term $P(\mathbf{u}^i)$ in {\eqref{eq: Bayes rule_MAP}} appears as the bias term for sequential decoding. 
Also, it is assumed that $P(\mathbf{u}^{i-1}|u_i) = P(\mathbf{u}^{i-1})$ when deriving {\eqref{eq: Listbitmetric}}, where $u_i$ is the output of the convolutional code in PAC codes.
For the outer convolutional code, to heuristically derive this metric, we use the conventional ensemble of random codes for linear codes, as described in \cite[p 206]{gallager1968information}.
This ensemble is denoted by the pair $(\mathbf{T},\mathbf{c})$ as
$\mathbf{u} = \mathbf{v}\mathbf{T} + \mathbf{c}$, where $\mathbf{T}$ is a fixed but arbitrary upper-triangular Toeplitz matrix and $\mathbf{c}$ is a fixed but arbitrary $N$-length offset binary vector. 
In this ensemble of codes, $P(\mathbf{u}^{i-1}|u_i) = P(\mathbf{u}^{i-1})$ holds.

The partial path metric up to the $i$th node would be
\begin{equation}
    \Gamma(\mathbf{u}^i;\mathbf{y}) = \sum_{j =1}^i \gamma_j(u_j;\textbf{y},{\textbf{u}}^{j-1}).
\end{equation}

At each step, if the bit is an information bit, the top-of-the-stack's path is replaced by its $2$ successors that have been extended by one branch, and the partial path metric is summed with the corresponding branch metrics.
If the bit is a frozen bit, the top-of-the-stack's path is replaced by its successor that has been extended by one branch, and the partial path metric is summed with the corresponding branch metric.
The algorithm continues in this manner until the decoding tree reaches a leaf node. 
For a message sequence of length $K$, in step $2$ of the algorithm, it is just necessary to branch the tree for the information bits.
For the frozen bits, it is necessary to compute only one new metric at step $2$. 
As a result, the tree code has $2^K$ branches for a PAC$(N, K)$ code. 
The stack size increases by one for each visited node for the information bits but remains unchanged for the frozen bits. 
The decoding algorithm may visit each level of the decoding tree many times.
It is proved that in convolutional codes, a sequential decoding algorithm visits each level of the decoding tree almost once for the rates below the channel cutoff rate (i.e., for high SNR values) \cite{gallager1968information}. 
When the code rate profile is less than the polarized channel cutoff rate profile, the findings indicate that the sequential decoding of PAC codes has a similar computational complexity  \cite{moradi2021sequential, moradi2021monte}.

\begin{example}
As a small instance, consider a PAC$(8,4)$ code using $\mathcal{A} = \{4, 6, 7, 8\}$ as a rate-profiling and the matrix $\mathbf{T}$ made using the generator polynomial $\mathbf{c}(t) = t^{7} + t^{6} + t^{4}+ 1$ (321 in octal notation).
Other numerical results in this paper use
$\mathbf{c}(t) = t^{10} + t^{9} + t^{7} + t^{3} + 1$ (3211 in octal notation \cite{moradi2020performance}).
Assume that the data vector is $\mathbf{d} = (1, 0, 0, 1)$ and
hence $\mathbf{v} = (0, 0, 0, 1, 0, 0, 0, 1)$ and $\mathbf{u} = (0, 0, 0, 1, 1, 0, 1, 1)$.
Assume that at $2.5$ SNR value, the receiving vector is $\mathbf{r}=(-1.68, -0.74, 1.71, -2.3, 1.07, 2.03, -1.69, 0.22)$.

\begin{figure}[htbp]
\centering
	\includegraphics [width = \columnwidth]{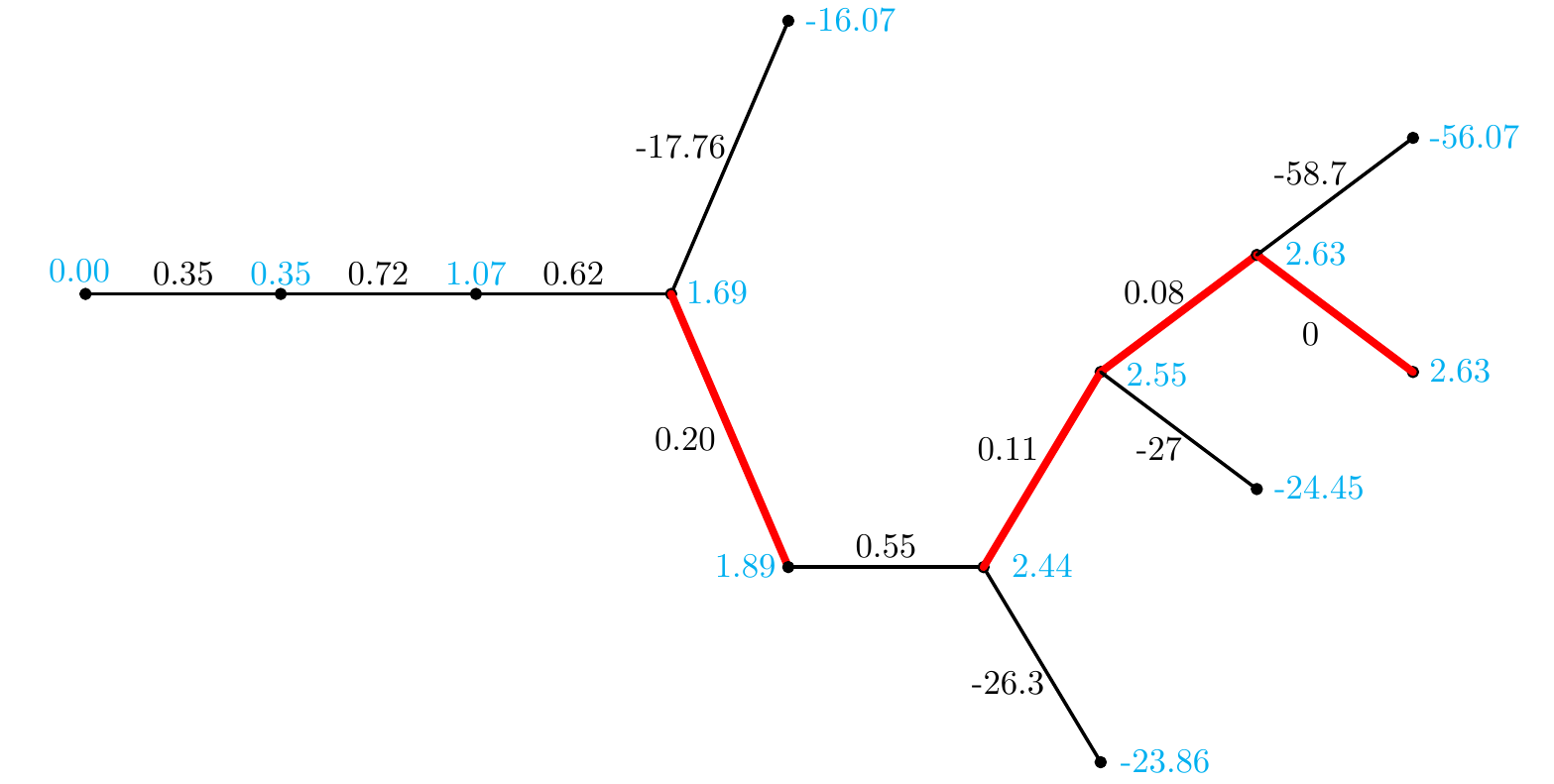}
	\caption{Explored binary tree by stack decoding algorithm.} 
	\label{fig: PACTree_stack}
\end{figure}

In Fig. \ref{fig: PACTree_stack}, the investigated part of the tree is shown with each branch of the tree labeled with its corresponding bit-metric value.
The partial path metrics values to a node are also shown with a blue color on top of that node.
At time 0, the decoder begins at the root and traverses the tree from left to right.
Whenever the corresponding information bit is 0, it selects the upper branch; when it is 1, it chooses the lower branch. 
Also, the contents of the stack are shown in Table \ref{table:PACstack} after each partial path metrics reordering. 
The stack maintains the input bits as well as the metrics associated with each path. 
The algorithm is ended after the eighth cycle.
Note that the stack size is not increased at steps corresponding to the frozen bits since, at these steps, the code tree does not branch for them. 
The algorithm's decision is the path $00010001$ or the information sequence $1001$. 
Bold Red branches in the code tree indicate selected information branches.
Notice that the stack decoding algorithm's significant challenges are storing many elements in the stack and the sorting operation at each level. 
The minimum possible elements in the stack is $K+1$.

\end{example}

\begin{table*}[htbp]
\scriptsize
\centering
\caption{The stack content and its partial path metric values after the ordering in each decoding iteration.}
\begin{tabular}{llllllllll}
\multicolumn{1}{c||}{Step} & 
\multicolumn{1}{c}{0} & 
\multicolumn{1}{c}{1} & 
\multicolumn{1}{c}{2} & 
\multicolumn{1}{c}{3} & 
\multicolumn{1}{c}{4} & 
\multicolumn{1}{c}{5} & 
\multicolumn{1}{c}{6} & 
\multicolumn{1}{c}{7} & 
\multicolumn{1}{c}{8}                              \\ \hline\hline
\multicolumn{1}{l||}{\begin{tabular}[c]{@{}l@{}}Stack \\ Content\end{tabular}} 
& -, (0.00) 
& 0 (0.35) 
& 00 (1.07) 
& 000 (1.69) 
& \begin{tabular}[c]{@{}l@{}}0001 (1.89)\\ 0000 (-16.07)\end{tabular} 
& \begin{tabular}[c]{@{}l@{}}00010 (2.44)\\ 0000 (-16.07)\end{tabular} 
& \begin{tabular}[c]{@{}l@{}}000100 (2.55)\\ 0000 (-16.07)\\ 000101 (-23.86)\end{tabular} 
& \begin{tabular}[c]{@{}l@{}}0001000 (2.63)\\ 0000 (-16.07)\\ 000101 (-23.86)\\ 0001001 (-24.45)\end{tabular} 
& \begin{tabular}[c]{@{}l@{}}00010001 (2.63)\\ 0000 (-16.07)\\ 000101 (-23.86)\\ 0001001 (-24.45)\\ 00010000 (-56.07)\end{tabular} \\ 
\end{tabular}\label{table:PACstack}
\end{table*}

\section{Improving stack algorithm}\label{sec: reliable_stack}
According to  \cite{moradi2021sequential}, the $i$th bit-channel metric function \eqref{BranchMetric} results in a sequential decoding with a low computational complexity.
Similar to what we previously described, in the stack algorithm, the bit metric of an incorrect branch in noiseless bit channels converges to $-\infty$. 
As a result of this finding, any pathways that include the wrong branch of a noiseless channel would never reach the top of the stack. 
Ideally, in the noiseless bit channel, there is no need to insert the path of the wrong branch into the stack at all, hence the number of elements in the stack would be kept very small.
It would be interesting to see whether we can take advantage of this in the case of short block length codes. 
We will put this to the test using a PAC$(128, 64)$ stack decoding. 
Assume that the bit-metric value of less than $m_T = -20$ (instead of $-\infty$) corresponds to the bit-metric value of a wrong branch.
For the bit channels with a bit-metric value smaller than $-20$, Table \ref{chX:table: wrongBranch} displays the levels of the decoding tree, the associated bit-channel cutoff rates, and the bit-channel metric values of both corresponding branches in a single decoding trial at $2.5$ dB SNR. 
This table has $54$ rows, which indicates that the stack size would not need to expand in this $54$ times in comparison to the conventional stack algorithm. 

\begin{table}[htbp]
\footnotesize
\centering
\caption{Levels, reliabilities, and bit-metric values of both branches (up and down branches) of noiseless bit channels at 2.5 dB SNR.}
\begin{tabular}{llll}
i   & $E_0(1,W_N^{(i)})$ & up      & down    \\ \hline
32 		& 0.9978 		& 0.00 		& -38.96 	 \\ \hline
46 		& 0.9367 		& 0.06 		& -22.94 	 \\ \hline
47 		& 0.9612 		& -25.00 		& 0.04 	 \\ \hline
48 		& 0.9997 		& -55.91 		& 0.00 	 \\ \hline
52 		& 0.9531 		& 0.05 		& -23.89 	 \\ \hline
54 		& 0.9786 		& 0.02 		& -33.29 	 \\ \hline
55 		& 0.9876 		& 0.01 		& -35.21 	 \\ \hline
56 		& 1.0000 		& 0.00 		& -73.56 	 \\ \hline
58 		& 0.9921 		& -32.25 		& 0.01 	 \\ \hline
59 		& 0.9954 		& -37.69 		& 0.00 	 \\ \hline
60 		& 1.0000 		& 0.00 		& -77.40 	 \\ \hline
61 		& 0.9975 		& -41.48 		& 0.00 	 \\ \hline
62 		& 1.0000 		& 0.00 		& -91.17 	 \\ \hline
63 		& 1.0000 		& 0.00 		& -Inf 	 \\ \hline
64 		& 1.0000 		& -191.64 		& 0.00 	 \\ \hline
78 		& 0.9904 		& 0.01 		& -22.10 	 \\ \hline
79 		& 0.9946 		& 0.01 		& -23.38 	 \\ \hline
80 		& 1.0000 		& 0.00 		& -59.10 	 \\ \hline
86 		& 0.9977 		& -34.53 		& 0.00 	 \\ \hline
87 		& 0.9988 		& 0.00 		& -32.41 	 \\ \hline
88 		& 1.0000 		& -79.83 		& 0.00 	 \\ \hline
90 		& 0.9992 		& -27.72 		& 0.00 	 \\ \hline
91 		& 0.9995 		& 0.00 		& -38.39 	 \\ \hline
92 		& 1.0000 		& 0.00 		& -79.27 	 \\ \hline
93 		& 0.9998 		& 0.00 		& -41.88 	 \\ \hline
94 		& 1.0000 		& -87.67 		& 0.00 	 \\ \hline
95 		& 1.0000 		& -Inf 		& 0.00 	 \\ \hline
96 		& 1.0000 		& 0.00 		& -209.46 	 \\ \hline
100 		& 0.9987 		& -25.97 		& 0.00 	 \\ \hline
102 		& 0.9995 		& -47.14 		& 0.00 	 \\ \hline
103 		& 0.9997 		& -46.93 		& 0.00 	 \\ \hline
104 		& 1.0000 		& -103.52 		& 0.00 	 \\ \hline
106 		& 0.9998 		& -36.09 		& 0.00 	 \\ \hline
107 		& 0.9999 		& 0.00 		& -28.38 	 \\ \hline
108 		& 1.0000 		& 0.00 		& -81.81 	 \\ \hline
109 		& 1.0000 		& -Inf 		& 0.00 	 \\ \hline
110 		& 1.0000 		& -Inf 		& 0.00 	 \\ \hline
111 		& 1.0000 		& 0.00 		& -Inf 	 \\ \hline
112 		& 1.0000 		& 0.00 		& -249.96 	 \\ \hline
114 		& 1.0000 		& -50.76 		& 0.00 	 \\ \hline
115 		& 1.0000 		& -45.28 		& 0.00 	 \\ \hline
116 		& 1.0000 		& -106.58 		& 0.00 	 \\ \hline
117 		& 1.0000 		& 0.00 		& -Inf 	 \\ \hline
118 		& 1.0000 		& -Inf 		& 0.00 	 \\ \hline
119 		& 1.0000 		& 0.00 		& -Inf 	 \\ \hline
120 		& 1.0000 		& 0.00 		& -283.56 	 \\ \hline
121 		& 0.9999 		& 0.00 		& -51.19 	 \\ \hline
122 		& 1.0000 		& 0.00 		& -Inf 	 \\ \hline
123 		& 1.0000 		& 0.00 		& -Inf 	 \\ \hline
124 		& 1.0000 		& -Inf 		& 0.00 	 \\ \hline
125 		& 1.0000 		& 0.00 		& -Inf 	 \\ \hline
126 		& 1.0000 		& -Inf 		& 0.00 	 \\ \hline
127 		& 1.0000 		& -Inf 		& 0.00 	 \\ \hline
128 		& 1.0000 		& 0.00 		& -641.35 	\\ \hline
\end{tabular}\label{chX:table: wrongBranch}
\end{table}

\begin{figure}[htbp]
\centering
	\includegraphics [width = \columnwidth]{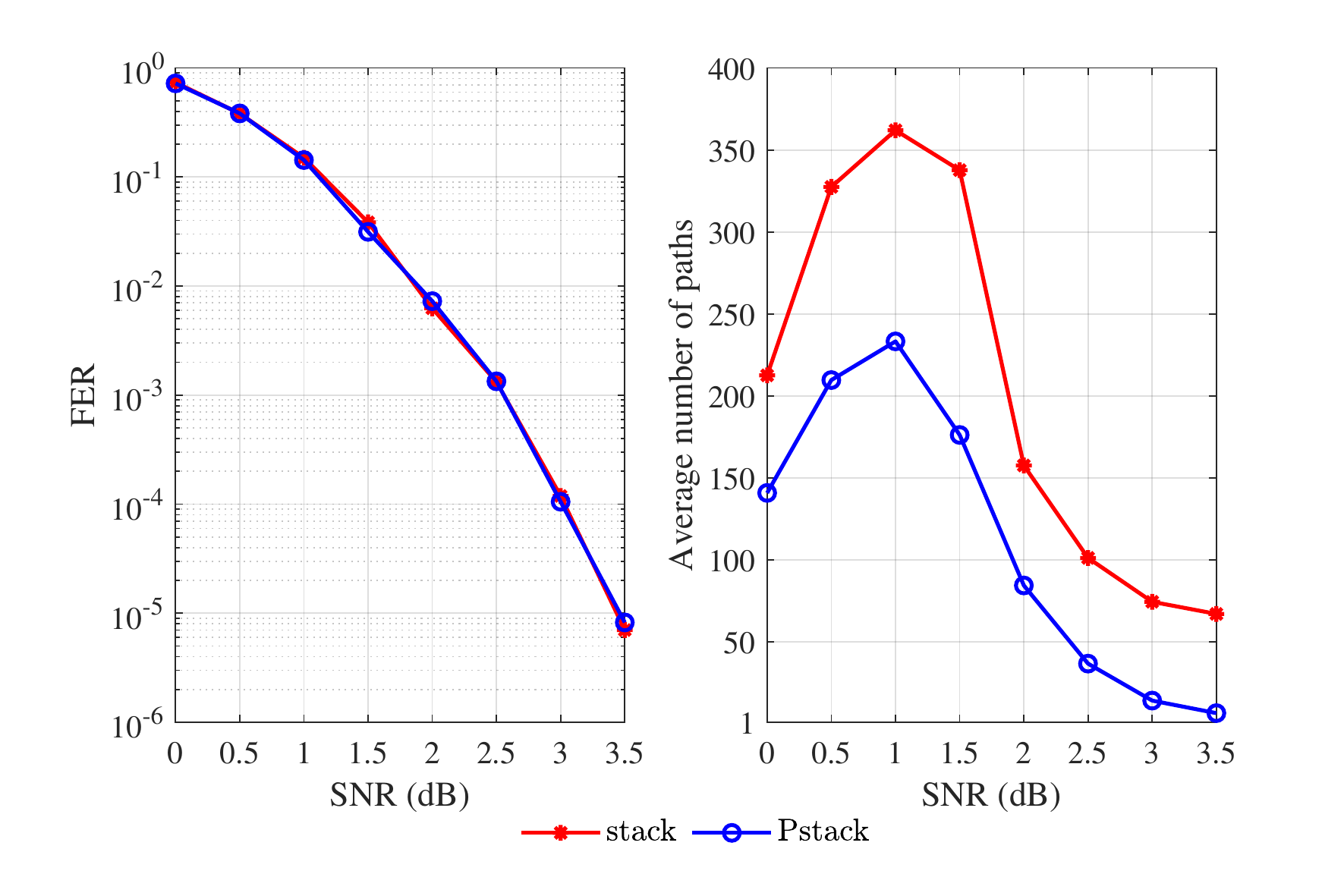}
	\caption{FER performance and average number of paths in the stack comparison of PAC$(128,64)$ codes with $m_T = -20$. } 
	\label{fig: prunned_ithbit}
\end{figure}

Figure \ref{fig: prunned_ithbit} compares the FER performance of our proposed stack algorithm (Pstack) as we do not insert the paths containing the wrong branches (a branch with the bit-metric value less than $m_T = -20$) with a conventional stack algorithm for a PAC$(128, 64)$ code.
As this figure shows, there is no loss in the error-correction performance.
This figure also compares the corresponding average number of elements in the stacks. 
Compared to the conventional stack algorithm, our proposed stack algorithm has much less elements in the stack.
The average number of elements in the stack is reduced from $67.04$ to $6.55$ at $3.5$ dB SNR. 

\section{Dynamic threshold}\label{sec: dynamic}
We proved that the probability of error due to pruning decreases exponentially with $m_T$.
If the FER performance at $3.5$ value is about $10^{-6}$ in the stack decoding with no pruning technique, it would be reasonable to use $m_T = -23$ ($2^{-23} \approx 10^{-7}$) at this SNR value.
In this manner, we may choose an appropriate value for $m_T$ for all other desired SNR values. 
The dispersion approximation yields a theoretical best attainable performance for every finite block-length code \cite{polyanskiy2010channel}.
If the FER performance achieved by the dispersion approximation for a particular code is $D$ at a specified SNR value, we use
\begin{equation}
    m_T = \lfloor \log_2(D/10) \rfloor
\end{equation}
as the dynamic bit-metric threshold for that SNR value. 
In this way, for a $(128, 64)$ code and SNR vector of $(0, 0.5, \cdots, 3.5)$, the corresponding threshold values are as
$$\mathbf{m}_T = (-5, -6, -7, -9, -11, -14, -18, -23).$$

\begin{figure}[htbp]
\centering
	\includegraphics [width = \columnwidth]{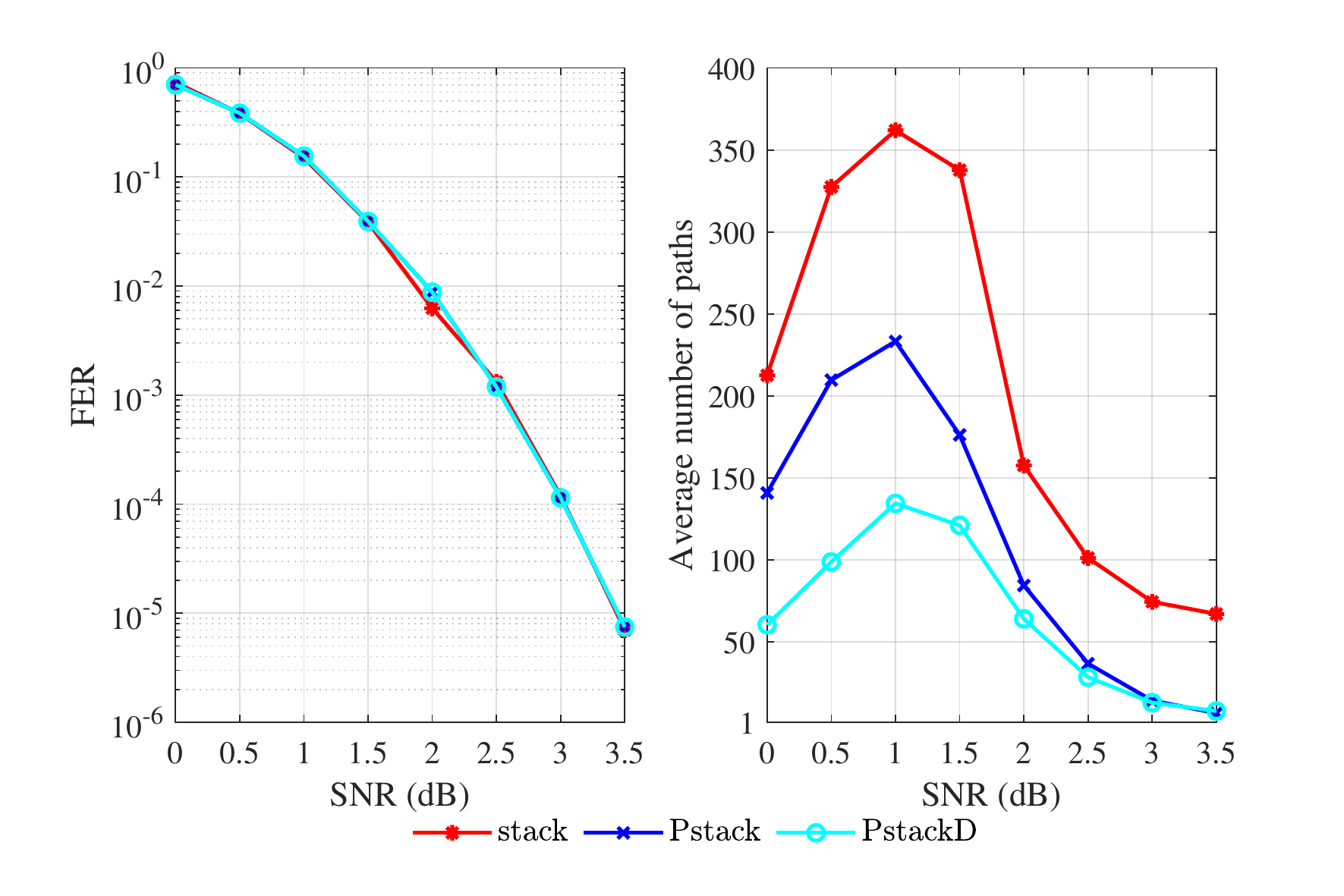}
	\caption{FER performance and average number of paths in the stack comparison of PAC$(128,64)$ codes with fixed and dynamic thresholds. } 
	\label{fig: PstackDynamic}
\end{figure}

Fig. \ref{fig: PstackDynamic} compares the FER performance and average number of elements in the stack of the conventional stack and Pstack ($m_T = -20$ for all SNR values) algorithms with our proposed algorithm when employing this dynamic
threshold settings (PstackD) for a PAC$(128, 64)$ code.
When compared to the Pstack approach, the PstackD technique results in a further reduction in the average number of elements in the stack for low SNR levels.  
For example, at a $1$ dB SNR value, the average number of elements in the stack of the conventional stack algorithm is $364$, the average number of elements in the stack of the Pstack approach is $233$, and the average number of elements in the stack of PstackD is around $134$. 

\section{Conclusion}\label{sec: conclusion}
In this paper, we provided and analyzed the optimal metric function of the SCL decoding algorithm for polar and PAC codes. 
On average, the path metric of the correct path should be equal to the sum of the bit-channel mutual information, and the bit metric of the incorrect branch can be no more than 0. 
We took advantage of this by introducing an approach to reject the erroneous paths based on the departure of their path metric from the bit-channel mutual information. 
This approach avoids sorting on many branches while causing no loss in the FER performance. 
Moreover, we proposed a similar approach to the stack algorithm based on the bit metrics of the polarized channels.
For the reliable bit channels, we proved that the bit-channel metric value should be almost zero for the correct branch and a very large negative number for the wrong branch. 
This way, the decoding algorithm can identify the incorrect branch and avoid adding it to the stack. 
This reduces the average number of elements in the stack of up to $90\%$ in our simulation results. 
Additionally, we proved that for a threshold value smaller than the bit-channel cutoff rate, the probability of pruning the correct path exponentially approaches zero, allowing us to suggest a technique for determining the threshold value.


\ifCLASSOPTIONcaptionsoff
  \newpage
\fi

\bibliographystyle{IEEEtran}
\bibliography{bibliography}

\end{document}